\documentclass[lettersize,journal]{IEEEtran}
\usepackage{amsmath,amsfonts}
\usepackage{algorithmic}
\usepackage{algorithm}
\usepackage{array}
\usepackage[caption=false,font=normalsize,labelfont=sf,textfont=sf]{subfig}
\usepackage{textcomp}
\usepackage{stfloats}
\usepackage{url}
\usepackage{verbatim}
\usepackage{graphicx}
\usepackage{cite}
\usepackage{booktabs}
\begin{document}

\title{Heterogeneous Layered Structures \\ Can Modulate Human Softness Perception}

\author{
Yuno Higuchi,
Yosuke Iwashita,
Yuji Ohgi,
Masashi Nakatani*

\thanks{*Corresponding authors: Masashi Nakatani (nakatani.z3@keio.jp). Y.N. and Y.O. are also with Graduate School of Media and Governance, Keio University, Japan. All authors are with Faculty of Environment and Information Studies, Keio University, Japan. 
}
\thanks{Manuscript received April XX, 2026; revised July XX, 2026.}
}

\markboth{Transatsions on Haptics,~Vol.~xx, No.~xx, July~202x}%
{Shell \MakeLowercase{\textit{et al.}}: A Sample Article Using IEEEtran.cls for IEEE Journals}


\maketitle

\begin{abstract}
Human softness perception in haptics has mainly been studied using mechanically homogeneous objects, despite the fact that many real-world objects exhibit heterogeneous layered structures with nonuniform stiffness. 
This study examined how layered heterogeneity modulates haptic softness perception. 
Sixteen lattice-structured stimuli were fabricated by 3D printing, with the stiffness of the upper four layers systematically varied while the bottom two layers remained fixed. 
Twenty-two participants evaluated the softness of the stimuli in a psychophysical task, and compression tests were conducted to quantify their mechanical properties. 
Perceived softness was significantly predicted by displacement under load, however, perceptual ranking did not fully coincide with the physical ranking. 
Linear mixed-effects analyses showed that the softness of the outermost layer had the greatest impact on the perceived softness. 
Perceived softness also increased as the number of soft subsurface layers increased, although this contribution decreased with depth.
Layers 2 and 3 showed significant effects, whereas Layer 4 did not. 
These findings suggest that haptic softness perception depends not only on the overall stiffness but also on the depth-dependent distribution of compliance within layered structures.
\end{abstract}


\section{Introduction}
\IEEEPARstart{T}{ouch} plays an indispensable role in accurately evaluating the physical properties of objects, such as shape, texture, hardness, and temperature. 
The sensory information obtained through touch provides perceptual experiences that cannot be differentiated from those obtained through other sensory modalities. 
In particular, detailed information about an object’s structure and material properties is often only apparent through active manual exploration. 
During active touch exploration, people unconsciously adapt their exploratory procedures according to the properties of an object, such as its texture and softness/hardness \cite{Lederman1987}.

Decades of haptic research have shown that the perception of an object’s softness is primarily understoodas a phenomenon that depends on proprioceptive information, especially force-related cues such as the relationship between force and displacement during pressing\cite{Srinivasan1995}\cite{Bergmann2009}. 
Other previous studies have suggested that the relationship between contact force and contact-area spread provides an important cue for haptic softness perception, and that changes in fingertip contact area can also contribute to the perception of finger displacement.\cite{Bicchi2000}\cite{Moscatelli2016}.

However, most of the major findings on softness perception in haptics have been obtained using objects with spatially uniform stiffness. 
In contrast, many objects that we touch have heterogeneous layered structures composed of multiple layers with different hardnesses and materials in everyday life. 
Human skin is an example of a mechanically heterogeneous structure, and palpation-based examination relies, at least in part, on detecting relatively hard masses embedded within softer surrounding tissue\cite{Klein2005}.
Although there has been an experimental study on softness perception in mechanically heterogeneous objects using plums \cite{Xu2020}, there are still only limited examples of how heterogeneity in object stiffness influences the perceived softness.

\begin{figure}[t]
    \centering
    \includegraphics[width=1.0\linewidth]{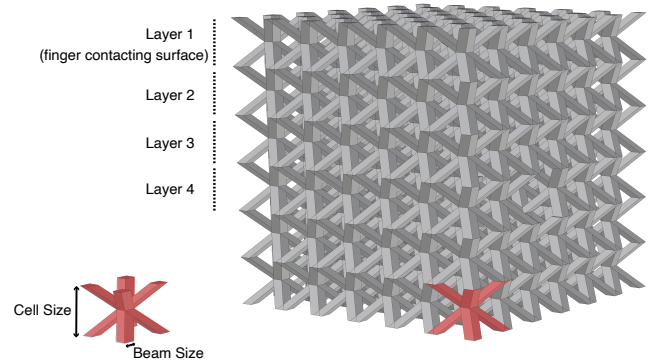}
    \caption{Unit cell (left) and six-layer lattice structure (right)}
    \label{rhino}
\end{figure}

In the present study, we fabricated heterogeneous layered structures with varying compliance using 3D printing technology and investigated how such heterogeneous layered structures affect human softness perception. 
Specifically, we designed a six-layer structure in which the stiffness of the top four layers was systematically varied by adjusting the beam thickness of each layer (Fig. \ref{rhino}).
We then tested the following two hypotheses regarding the softness perception of heterogeneous layered structures:
\begin{itemize}
  \item Hypothesis 1: Stimuli with a soft surface layer are perceived as softer than those stimuli with a hard surface layer, regardless of the properties of the underlying layers.
  \item Hypothesis 2: Even for stimuli with a hard surface layer, the perceived overall softness increases as the number of soft layers beneath the surface increases.
 
\end{itemize}

Part of the data used in this study has been published in a book chapter as a brief review \cite{Nakatani2026}, however, the present study offers additional discussion of their perceptual interpretation, particularly from the perspective of haptic softness.

\section{Lattice Structures Used in the Present Experiment}

A lattice structure is a three-dimensional structure formed by the repeated arrangement of unit cells (Fig. \ref{rhino}). 
In this study, the cell size was fixed at 6 mm, and the softness of the structure was controlled by varying the thickness of the struts (i.e., beam size). 
Lattice structures were designed using the 3D CAD software Rhinoceros (Robert McNeel \& Associates(TLM, Inc.), Seattle, USA) and its Grasshopper plugin.

The experimental stimuli were fabricated using a fused deposition modeling (FDM) 3D printer (Bambu Lab P1S), and a TPU 95A filament (SainSmart). 
Each stimulus consisted of a lattice-structured block measuring 3.6 $\times$ 3.6 $\times$ 3.6 [cm]. 
To examine the effects of structural differences, the bottom two layers were fixed as hard layers in all stimuli, while the overall stiffness was manipulated by varying the combinations of hard and soft layers in the upper four layers.

The beam size of the hard layers was set as 1.0 mm, whereas that of the soft layers was set as 0.75 mm. 
To determine the beam sizes that would produce perceptible, but not excessively large differences in haptic softness, we first examined the manufacturing limits of the printer. 
The results showed that 0.75 mm was the thinnest beam size that could be printed reliably and cleanly with the printer used in this study. 
In contrast, the beam size for the hard layers was set to 1.0 mm, which was the maximum thickness that could be pressed comfortably without causing pain to the finger. 
Based on these considerations, two beam-size conditions were established: 1.0 mm for hard layers and 0.75 mm for soft layers.

\begin{figure}[!t]
    \centering
    \includegraphics[width=0.85\linewidth]{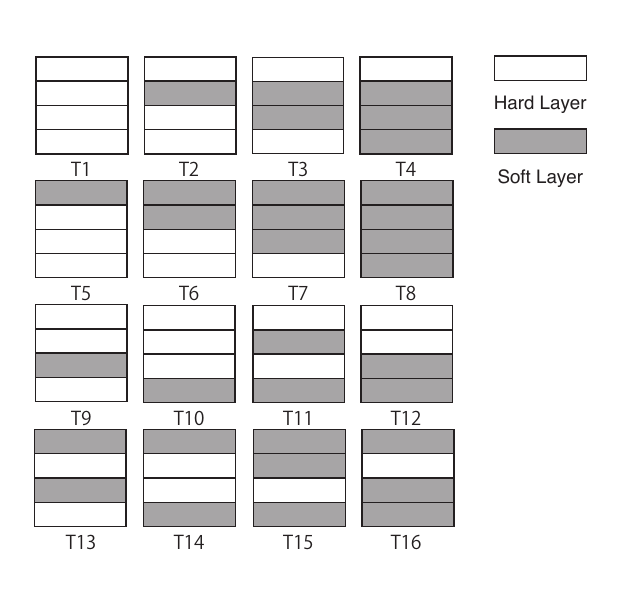}
    \caption{Overview of the sixteen stimuli used in the experiment and their layer combinations. Hard layers consist of a thick beam (1.0 [mm]) while soft layers consist of a thin beam (0.75 [mm]).}
    \label{fig:combination}
\end{figure}

Sixteen stimuli (T1--T16) were prepared, and their layer combinations are shown in Fig. \ref{fig:combination}. 
To test Hypotheses 1 and 2, T1--T8 were designed by manipulating the stiffness of the upper four layers. 
T9--T16 were additionally prepared to extend the degree of heterogeneity relative to T1--T8 and to examine its influence on softness perception in greater detail.

\section{Experiment}

\subsection{Mechanical characterization of the stimuli}
Compression tests were conducted using a tabletop universal testing machine to quantify the physical stiffness of experimental stimuli. 
Specifically, a force tester (MCT-2150W, A\&D Co., Ltd. Tokyo, Japan) was used with an attached compression plate, and the maximum load was set to 500 [N]. 
The displacement at 3000 [gf] was extracted from the force–displacement data, as participants in the psychological evaluation described below were instructed that they could apply up to 3000 [gf] to each stimulus.
Because samples fabricated by 3D printing may show run-to-run variability, five sets of all sixteen stimulus types were prepared for mechanical characterization.


\subsection{Psychological evaluation of the stimuli}

\subsubsection{Participants}
Twenty-two university students aged 21 -- 24 years old participated in the experiment (13 women and 9 men; mean age = 21.8 years).
All the participants were right-handed. 
The experimental protocol was approved by the Research Ethics Committee of Keio University Shonan Fujisawa Campus (approval no. 586).

\subsubsection{Procedure}
In the psychophysical experiment, participants evaluated the softness of each stimulus using the index finger of their dominant hand. 
The softness of the reference stimulus (T11; see Fig. \ref{fig:combination}) was set to 50.
Participants were instructed to assign a value greater than 50 if a stimulus felt softer than the reference and a value less than 50 if it felt harder than the reference, with responses restricted to values greater than 1. 
The reference stimulus was selected because its physical stiffness was intermediate among the 16 stimuli, based on the results of the compression test (see Fig. \ref{fig:displacement}).

The presentation order of the stimuli was randomized across the participants. 
Because softness perception can be influenced by the contact area, applied force, and visual information, these factors were controlled as much as possible. 
Specifically, the force applied by the participant was limited to 3000 [gf].
Although the participants could visually confirm the position of the stimulus through a half mirror, they were unable to observe the extent of its deformation in detail. 
In addition, masking tape was attached to both lateral edges of the contact surface to restrict the size of contact area. 
No limit was imposed on the number of presses allowed in each trial.

\subsubsection{Data analysis}
For each stimulus, the softness ratings obtained from the participants were geometrically averaged across the 22 participants and normalized across participants. 
In addition, a multiple regression analysis was conducted to examine the relationship between the displacement produced under a 3000 [gf] load and the estimated softness ratings.

To further test Hypotheses 1 and 2, linear mixed-effects models (LMMs) were constructed and compared. 
\begin{itemize}
  \item To test Hypothesis 1, namely that the softness of the outermost layer in contact with the finger pad influences perceived softness, a model tha  included the softness of the contact surface (Model 1) was compared with a model that did not include this factor (Model 0).
  \item To test Hypothesis 2, that perceived softness increases as the number of soft layers increases, a model incorporating the number of soft underlying layers (Model 2) was constructed and compared with Model 1.
  \item In addition, we examined whether the contribution of soft layers to perceived softness varied as a function of the depth position (Model 3). 
  Defining the outermost layer in contact with the finger pad as Layer 1, we investigated up to which layer stiffness contributed to human softness perception.
\end{itemize}

Model comparisons were conducted using the likelihood ratio test (LRT) with the Akaike information criterion (AIC) used as the index for model evaluation. 
All linear mixed-effect analyses were performed using R (version 4.4.1).

\section{Results}

\subsection{Mechanical Characterization}
\begin{figure}[!t]
    \centering
    \includegraphics[width=1.0\linewidth]{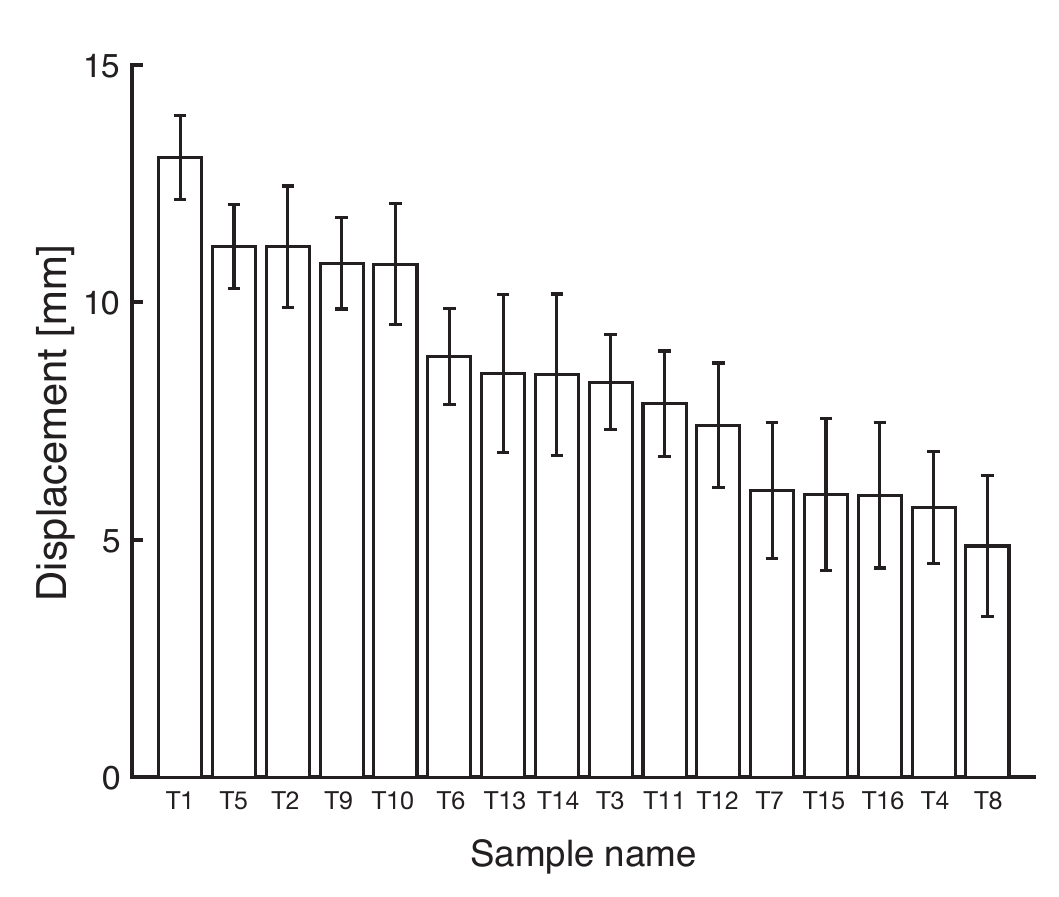}
    \caption{The displacement of each stimulus measured during compression test a load of 3000 [gf], measured during compression test. Error bars indicate standard deviations. Figure reprinted from Nakatani et al. (2026) with permission \cite{Nakatani2026}}
    \label{fig:displacement}
\end{figure}

The displacement of each stimulus measured in the compression test is shown in Fig. \ref{fig:displacement}. 
In the psychophysical evaluation, participants were allowed to press each stimulus with their fingertip up to a maximum load of 3000 [gf]. 
Therefore, the displacement values shown here correspond to those measured under a load of 3000 [gf].

\begin{figure*}[!t]
    \centering
    \includegraphics[width=0.98\textwidth]{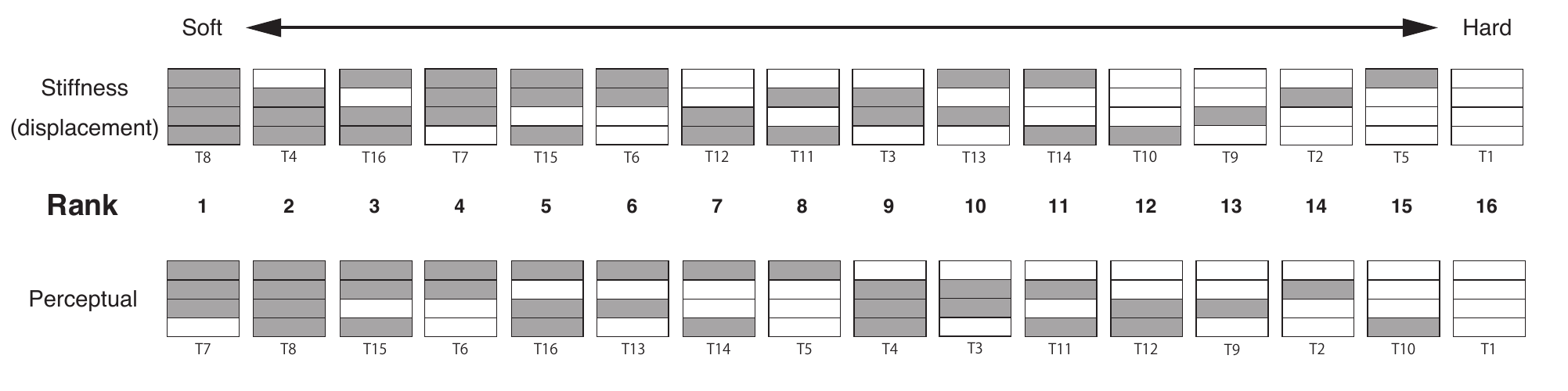}
    \caption{Comparison between the physical stiffness ranking obtained from the compression test and the perceptual ranking obtained from the softness evaluation task.}
    \label{fig:rank}
\end{figure*}

Fig. \ref{fig:rank} compares the rank order of physical softness obtained from the compression test with the perceptual rank order obtained from the softness evaluation task.
The figure shows that the physical and perceptual rankings did not completely coincide, indicating a discrepancy between the measured mechanical stiffness and the perceived softness.

\subsection{Psychological evaluation}
\begin{figure}[!t]
    \centering
    \includegraphics[width=1.0\linewidth]{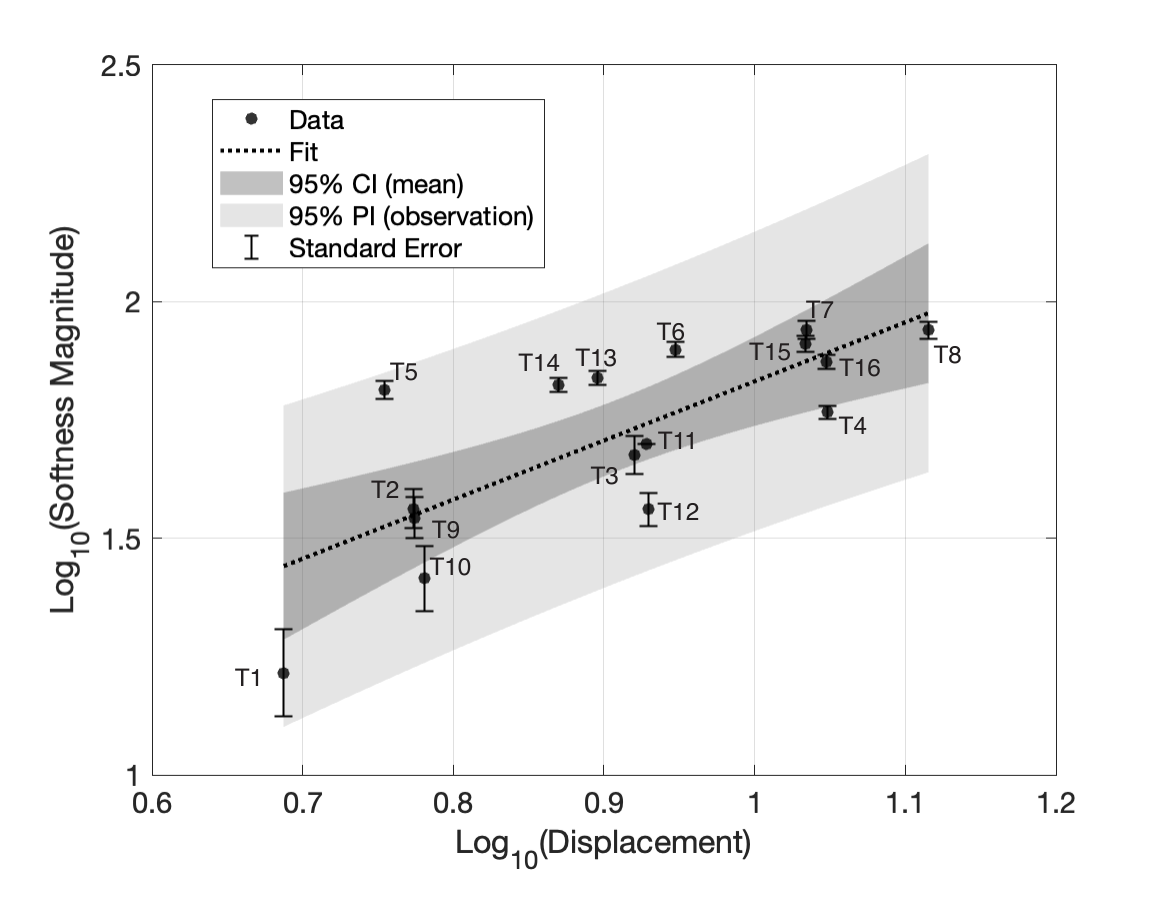}
    \caption{Relationship between displacement under a 3000 [gf] load and the estimated softness of each stimulus by the participants. CI: confidence interval; PI: prediction interval. Figure modified from Nakatani et al. (2026) with permission \cite{Nakatani2026}}
    \label{fig:plot}
\end{figure}

To examine the relationship between perceived softness and the physical properties of the stimuli, we fitted a linear regression model using the logarithm of displacement as the predictor and the logarithm of perceived softness as the response variable. 
The model was statistically significant, $F(1, 14) = 19.1$, $p < .001$, with an adjusted $R^2$ of 0.55 (Fig. \ref{fig:plot}). 
This result indicates that approximately half of the variance in perceived softness can be explained by the displacement of stimuli. 
In other words, our result suggested that the displacement under the compression test is an important predictor of softness perception. 
At the same time, although several stimuli fell within the 95\% prediction interval, part of the variance in perceived softness remained unexplained by the regression model.

Next we tested the hypothesis based on linear mixed-effects models. 
A comparison between Model 0 and Model 1 showed that Model 1 provided a significantly better fit to the data, $\chi^2(1) = 132.56$, $p < .001$. 
This result supports Hypothesis 1, namely that a soft outermost layer in contact with the finger pad enhances the perceived softness. 
The marginal $R^2$, representing the proportion of variance explained by the fixed effects, was 0.51.

To test Hypothesis 2, Model 2, which incorporates the number of soft layers, was compared with Model 1. 
Model 2 showed a better fit to the data ($\Delta AIC = -86.9$, $p < .001$), and the marginal $R^2$ increased by 0.17. 
This result suggests that the perceived softness increases as the number of soft layers increases.

\begin{figure}[!t]
    \centering
    \includegraphics[width=0.8\linewidth]{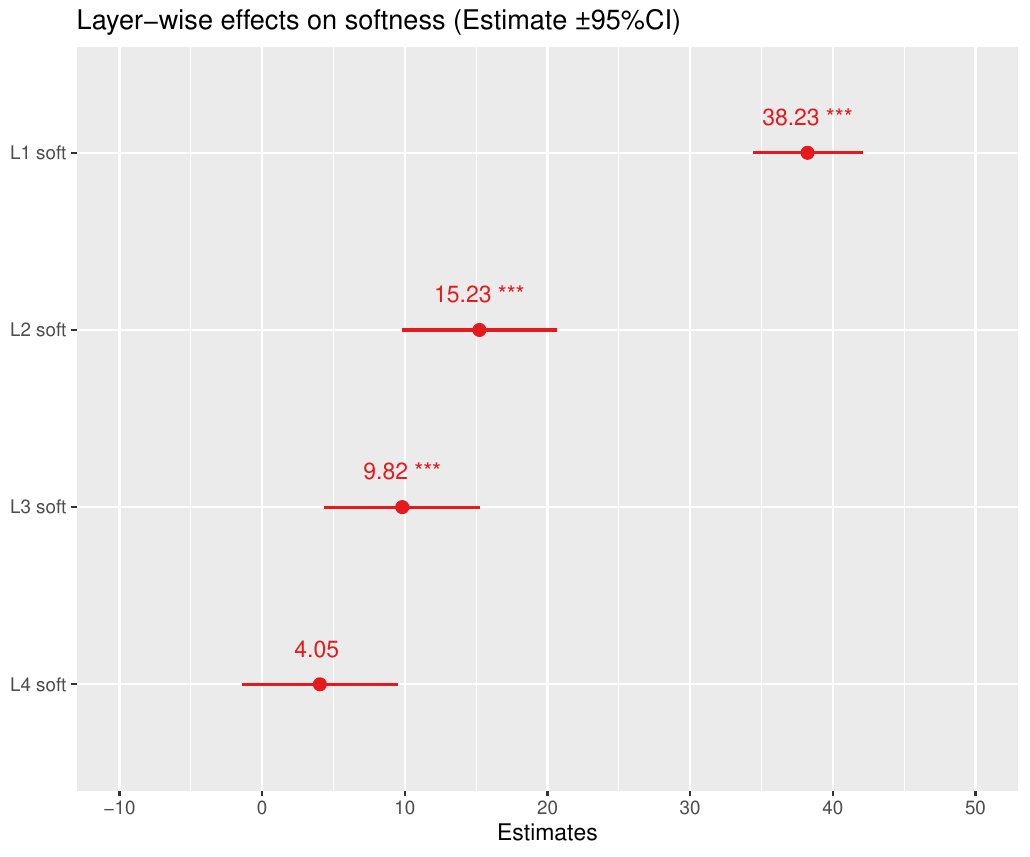}
    \caption{Effects of layer position from the surface on perceived softness. Error bars indicate 95\% confidence intervals.}
    \label{fig:layer_effects}
\end{figure}

Finally, Model 3, which incorporates the softness of each layer separately, showed a better fit than Model 2 ($\Delta AIC = -4.8$, $\chi^2(2) = 8.19$, $p < .02$). 
This model can be interpreted as examining how the contribution of layer-specific softness varies as a function of depth, with the outermost layer in contact with the finger pad defined as Layer 1. 
Consistent with the result of Model 1, the effect of Layer 1 was the largest among the four layers in which a soft lattice structure was present ($\beta = 38.23$, $SE = 1.95$, $p < .001$). 
The effects of Layer 2 and Layer 3 were also statistically significant, indicating that both layers contributed to perceived softness (Layer 2: $\beta = 15.23$, $SE = 2.76$, $p < .001$; Layer 3: $\beta = 9.82$, $SE = 2.76$, $p < .001$). 
In contrast, the effect of Layer 4 was not statistically significant, suggesting that its contribution to perceived softness was minimal ($\beta = 4.05$, $SE = 2.76$, $p = .15$). 
Fig. \ref{fig:layer_effects} summarizes the contribution of each layer to the perceived softness.

\section{Discussion}
The results supported Hypothesis 1. Specifically, perceived softness increased when the outermost layer consisted of a soft lattice structure. 
This finding provides a coherent account of the perceptual ranking shown in Fig. \ref{fig:rank}. 
In other words, stimuli with a soft outermost layer were consistently judged to be softer regardless of their overall physical stiffness, indicating that the softness of the surface layer makes a substantial contribution to perceived softness.

Hypothesis 2 was partially supported. Incorporating the number of underlying soft layers improved the prediction of perceived softness, suggesting that perceived softness tended to increase as the number of soft subsurface layers increased.

Further analysis of the contribution of individual layers revealed that Layer 2 made a significant positive contribution to perceived softness (Fig. \ref{fig:layer_effects}). 
Layer 3 also contributed significantly, although the estimated regression coefficient $\beta$ decreased with increasing depth. 
In contrast, the effect of Layer 4 was not statistically significant, regardless of whether the layer was hard or soft.

Taken together, our findings suggest that, although soft layers within a lattice structure do influence perceived softness, their effect decreases as a function of depth from the surface. 
In other words, the contribution of physical softness to perception appears to decay rapidly with increasing distance from the contact surface. This pattern suggests that the human haptic system processes softness information in a depth-dependent manner, placing greater weight on the mechanical properties of the shallower layers.

Xu et al. investigated mechanically heterogeneous natural objects (plums) and proposed that softness perception depends primarily on \textit{virtual stiffness}, derived from the relationship between force and displacement during active exploration, rather than on the contact area \cite{Xu2020}. 
They also reported behavioral evidence suggesting that humans actively seek reaction-force cues that support stiffness discrimination during exploration. 
In the present study, however, the highest perceived softness was observed when the outermost layer was soft. 
This finding suggests that local cutaneous cues, such as those related to the contact area, may play an important role in the softness perception of layered structures, even if such cues were not clearly identified in Xu et al.'s study of natural objects. 
In this respect, the present findings are also consistent with previous studies that showed that increases in contact area enhance perceived softness through cutaneous mechanisms \cite{Bicchi2000, Moscatelli2016}.

\begin{figure}[t!]
    \centering
    \includegraphics[width=1.0\linewidth]{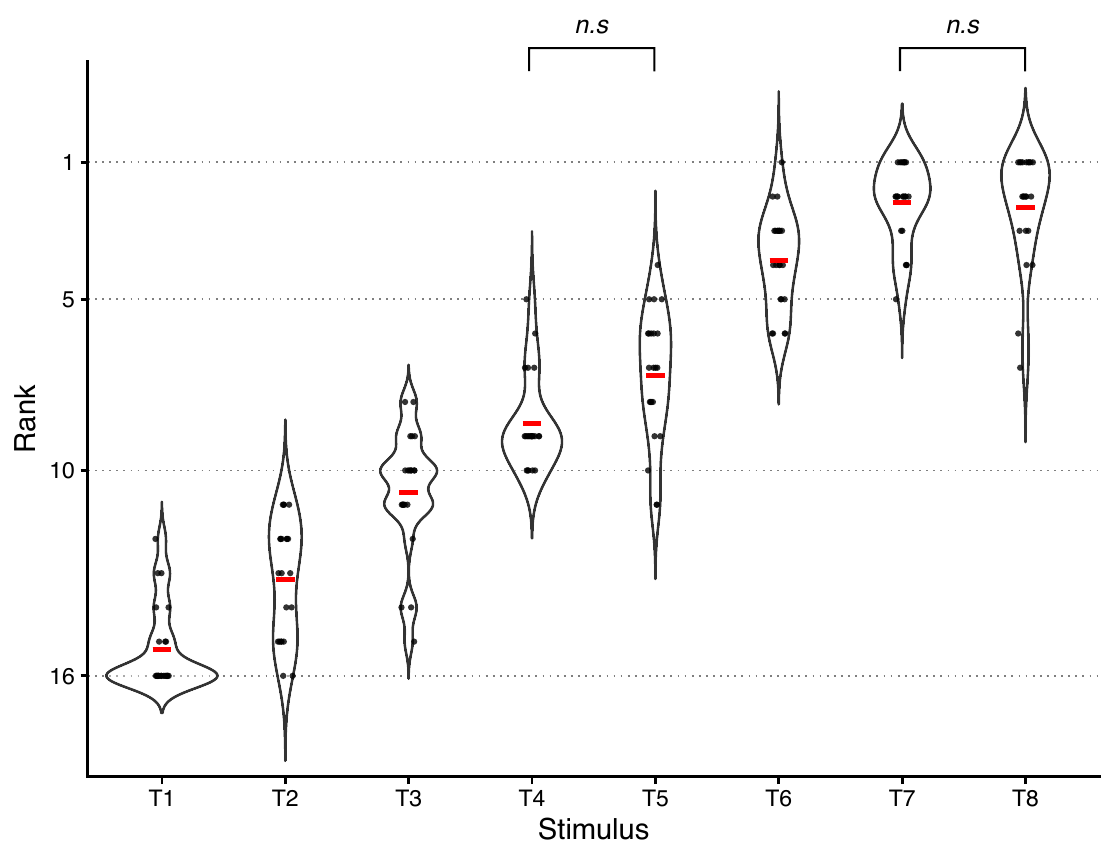}
    \caption{Softness ranking responses to experimental stimuli T1--T8 in the psychological evaluation. Higher ranks indicate that the stimulus was judged to be softer. 
    }
     \label{fig:rankAddAna}
\end{figure}

In the task in which the sixteen stimuli were ranked according to perceived softness, all of the top eight stimuli had a soft outermost layer (Fig. \ref{fig:rankAddAna}). 
This pattern may also be interpreted in relation to the phenomenon reported by Inoue et al., in which rigid plastic or rubber tactile stimuli containing a fingertip-sized concavity were perceived as softer than flat or convex stimuli \cite{Inoue2022}. 
In the present study, when the outermost layer was soft, the stimulus deformed readily to conform to the shape of the finger pad during pressing, resulting in a concave deformation. Consequently, the contact area between the finger pad and the stimulus likely increased, thereby enhancing perceived softness. 
The fact that stimuli T5, T13, and T14 exhibited greater perceived softness than predicted by the regression line may reflect an additional contribution of cutaneous cues to softness perception.

Based on this result, we conducted an additional analysis restricted to T1--T8.
Using T1--T8 stimuli, we performed a simple linear regression analysis with rank as the dependent variable and stimulus number as the independent variable. 
The analysis revealed that the stimulus number had a significant negative effect on rank ($\beta = -1.98$, $\mathrm{SE} = 0.055$, $t(174) = -35.99$, $p < .001$).
Specifically, for each one-unit increase in stimulus number, the mean rank decreased by approximately two points, indicating that the stimulus was perceived as higher (i.e., softer, Fig. \ref{fig:rankAddAna}).


A Friedman test was also conducted to examine the differences in rank among T1--T8. 
The result showed a significant effect of stimulus, $\chi^2(7, N = 22) = 140.0$, $p < .001$, Kendall's $W = .910$, indicating a large effect size.
Post hoc pairwise comparisons using Wilcoxon signed-rank tests with Holm correction revealed that the differences between T4 and T5 and between T7 and T8 were not statistically significant (Fig. \ref{fig:rankAddAna}).
In contrast, all other adjacent pairs showed statistically significant differences in softness rank (Table \ref{tab:posthoc_adjacent}).

\begin{table}[t!]
\centering
\caption{Post hoc pairwise comparisons for adjacent stimuli (Wilcoxon signed-rank tests).$p_{\mathrm{adj}}$ indicates the $p$ values adjusted for multiple comparisons using the Holm method.}
\label{tab:posthoc_adjacent}
\begin{tabular}{lrrrrl}
\toprule
Comparison & $n$ & $V$ & $p-value$ & $p_{\mathrm{adj}}$ & Significance \\
\midrule
T1 vs. T2 & 22 & 216 & .004 & .011 & * \\
T2 vs. T3 & 22 & 236 & $< .001$ & .002 & ** \\
T3 vs. T4 & 22 & 241 & $< .001$ & .001 & ** \\
T4 vs. T5 & 22 & 184 & .062 & .125 & n.s. \\
T5 vs. T6 & 22 & 253 & $< .001$ & $< .001$ & *** \\
T6 vs. T7 & 22 & 222 & .002 & .007 & ** \\
T7 vs. T8 & 22 & 120 & .840 & .840 & n.s. \\
\bottomrule
\end{tabular}
\end{table}


\subsection{Affect of heterogeneous layerd structures on softness perception}

The present study demonstrated that the perception of human softness is modulated by heterogeneous layered structures.  
In particular, our results supported Hypothesis 1, showing that perceived softness increases when the outermost layer consists of a soft lattice structure. 
This finding provides a coherent account of the perceptual ranking shown in Fig. \ref{fig:rank}. 
In other words, stimuli with a soft outermost layer were consistently judged to be softer regardless of their overall physical stiffness, indicating that the softness of the surface layer makes a substantial contribution to perceived softness.

Hypothesis 2 was partially supported. 
Incorporating the number of underlying soft layers improved the prediction of perceived softness, suggesting that perceived softness tended to increase as the number of soft subsurface layers increased. 
Further analysis of the contribution of individual layers revealed that Layer 2 made a significant positive contribution to perceived softness, and Layer 3 also contributed significantly, although the magnitude of the effect decreased with increasing depth (Fig. \ref{fig:layer_effects}). 
In contrast, the effect of Layer 4 was not statistically significant.
This was consistent with the result that there was not statistical significance between T7 and T8 in their rankings in pairwise comparisons (Fig. \ref{fig:rankAddAna}).
Taken together, our findings suggest that, although soft layers within a lattice structure influence perceived softness, their contribution decays rapidly with distance from the contact surface. 
This observation suggests that the human haptic system processes softness information in a depth-dependent manner, placing greater weight on the mechanical properties of shallower layers from contact surface.

These findings can be discussed in relation to those of previous studies on haptic softness perception. 
Xu et al. investigated mechanically heterogeneous natural objects (plums) and proposed that softness perception depends primarily on \textit{virtual stiffness}, derived from the relationship between force and displacement during active exploration, rather than on contact area \cite{Xu2020}. 
Xu et al. also reported behavioral evidence suggesting that humans actively seek force cues that support stiffness discrimination during exploration. 
However, in our study, the highest perceived softness was observed when the outermost layer was soft. 
Our results suggest that local cutaneous cues, such as those related to contact area, may play an important role in the softness perception of layered structures, even if such cues were not clearly identified in Xu et al.'s study of natural objects. 
In this respect, the present findings are also consistent with those of Bicchi et al., who showed that increases in contact area enhance perceived softness through cutaneous mechanisms \cite{Bicchi2000}\cite{Moscatelli2016}.

In the task in which the sixteen stimuli were ranked according to perceived softness, all the top eight stimuli had a soft outermost layer (Fig. \ref{fig:rank}). 
This pattern may also be interpreted in relation to the phenomenon reported by Inoue et al., in which rigid plastic or rubber tactile stimuli containing a fingertip-sized concavity were perceived as softer than flat or convex stimuli \cite{Inoue2022}. 
In the present study, when the outermost layer was soft, the stimulus deformed readily to conform to the shape of the finger pad during pressing, resulting in a concave deformation. 
Consequently, the contact area between the finger pad and the stimulus likely increased, thereby enhancing perceived softness. The fact that stimuli T5, T6, T13, and T14 exhibited greater perceived softness than predicted by the regression line may reflect an additional contribution of cutaneous cues to softness perception (Fig. \ref{fig:plot}).

\subsection{Limitations and Future Directions}
The present study has several limitations. 
One limitation of the present study is that the depth of pressing may have influenced how participants evaluated softness in the heterogeneous layered structures. 
In the experiment, participants were allowed to press the stimuli up to 3 kgf. However, some participants may not have pressed the stimuli deeply enough to substantially deform the subsurface layers, and may instead have based their judgments primarily on the softness of the surface layer. 
This possibility may explain the relatively large inter-participant variability observed for some stimuli, such as T2 and T5 (Fig. \ref{fig:rankAddAna}).
Accordingly, the absence of a significant contribution from Layer 4 may not be interpreted as evidence that deeper layers are irrelevant to softness perception in general. 

This issue is particularly intriguing when comparing T4 with T5. 
T4 had a hard surface layer with three soft layers underneath, whereas T5 had a soft surface layer with three hard layers underneath. 
Despite this structural contrast, no significant difference in softness rank was observed between these two stimuli. 
One possible interpretation is that participants differed in the extent to which they relied on surface softness versus subsurface stiffness. 
Consistent with this interpretation, the violin plot for T5 showed relatively large variation across participants.

Our findings suggest that the pressing depth is an important factor in the perception of softness for heterogeneous layered structures. 
Therefore, future studies should examine the relationship between indentation depth and softness judgments more directly, for example, by controlling the amount of displacement or applied force during active touch (See Xu et al. for their experiment \cite{Xu2020}). 
This approach would clarify the integration of haptic information from the surface and subsurface layers in human perception of softness.


Another important direction for future research is to investigate how finely humans can discriminate the  differences in the spatial distribution of softness within structurally heterogeneous objects. 
One advantage of lattice-based material design is that the softness of each layer can be independently manipulated. 
This makes it possible to ask not only whether layered heterogeneity affects perceived softness, but also to what extent humans can detect or discriminate differences in internal stiffness distributions. 
Addressing this question can contribute to the development of quantitative evaluation methods for research domains in which haptic assessment traditionally relies on subjective experience and/or hedonic judgment.


Surprisingly, humans were able to discriminate such subtle differences in softness (Fig. \ref{fig:rankAddAna}, Table \ref{tab:posthoc_adjacent}). 
Clarifying the cues underlying this ability will likely require future analyzes based on computational modeling, including models that consider the spatial distribution of mechanoreceptors in the skin. 
Previous studies have explained aspects of tactile shape perception based on the distribution of rapidly adapting (RA) afferents \cite{Nakatani2021a}. 
In future work, we aim to account for the human ability to discriminate slight differences in heterogeneous softness by combining skin deformation models, including finite element modeling (FEM), with neural firing models \cite{Ishizuka2025}.

Our study also highlights the practical value of fabricating haptic stimuli using 3D printing. 
3D printing offers the advantage of enabling flexible customization of products according to individual user preferences \cite{Nakatani2021, Uesaki2023}. 
As knowledge accumulates regarding the perceived softness generated by layered structures and the affective value associated with mechanical structures, we may be able to estimate an individual’s preferred softness  or haptic feeling and to translate that preference into a layered lattice design for digital fabrication using a 3D printer. 
In this sense, the present work provides a basis not only for understanding the perceptual mechanisms underlying softness perception in haptics, but also for future applications in the personalized design of  products that can be sensed by touch modality.

\section{Conclusion}

This study showed that heterogeneous layered structures affect how humans perceive softness.
In particular, the softness of the outermost layer had the strongest influence on the perceived softness, whereas the contribution of deeper layers progressively decreased with depth. 
Our findings suggest that human softness perception is shaped not only by the overall mechanical properties of an object but also by the depth-dependent distribution of stiffness within the structure. 
The  results provide a basis for designing multilayer haptic stimuli whose perceived softness can be controlled by structural configuration.

\section*{Acknowledgments}
This work was supported by JSPS KAKENHI grants JP20H05960 and 24K22311, as well as Taikichiro Mori Memorial research grants (2025, C-62) awarded to YH.
We would like to thank the Yasuaki Kakehi Lab at the University of Tokyo and Tetsuya Nagashima for their contributions to the initial conception of this study. We would also like to thank the members of the Yuji Ohgi Lab and the Nakatani Lab at Keio University for their valuable input.

\bibliographystyle{IEEEtran}
\bibliography{references}

\newpage

\end{document}